\documentclass{article}
\usepackage[utf8x]{inputenc}
\usepackage{subfig}

\usepackage[dvips]{graphicx}
\usepackage[english]{babel}
\usepackage{verbatim}
\usepackage{color}
\usepackage{algorithmic}
\usepackage{algorithm}
\usepackage{hyperref}
\usepackage{multirow}
\usepackage{epsfig}
\hypersetup{
    bookmarks=true,         
    unicode=true,          
    pdftoolbar=true,        
    pdfmenubar=true,        
    pdffitwindow=false,     
    pdfstartview={FitH},    
    pdftitle={How important tasks are performed: peer review},    
    pdfauthor={T. Hartonen,
    and M.J. Alava},     
    pdfsubject={},   
    pdfcreator={},   
    pdfkeywords={human behavior} {peer review} {decision making}, 
    pdfnewwindow=false,      
    colorlinks=true,       
    linkcolor=black,          
    citecolor=black,        
    filecolor=magenta,      
    urlcolor=cyan           
}
\title{How important tasks are performed: peer review}

\author{ T. Hartonen$^1$  and  M.J. Alava$^1$}

\begin{document}

\maketitle

\begin{abstract} The advancement of
various fields of science depends on the actions of individual
scientists via the peer review process. The referees' work patterns and stochastic nature of decision making both relate to the particular features of
refereeing and to the universal aspects of human behavior. Here, we
show that the time a referee takes to write a report on a scientific
manuscript depends on the final verdict. The data is compared to a
model, where the review takes place in an ongoing competition of
completing an important composite task with a large number of
concurrent ones - a Deadline -effect. In peer review human decision
making and task completion combine both long-range predictability
and stochastic variation due to a large degree of ever-changing
external "friction".
\end{abstract}

Peer review is one of the cornerstones of modern scientific
practice. Though various fields of science exhibit a wide variety of ways
this is implemented, they generally use the same central idea: anonymous scientists consider a manuscript, and then each in charge writes
a referee reply summarizing
the merits and weaknesses. The editor of the scientific journal acts
then. Sometimes the manuscript is  published directly,
sometimes after further improvement and review, and sometimes it is
rejected.

Currently the referee process is under a constant flux due e.g. to the
concomitant development of electronic(-only) scholarly journals
\cite{azar2007}. The publication practices of scientific journals
might need to be reshaped due to such pressures \cite{boldt2011}. It
is a long-standing question as to how such refereeing practices can guarantee a fair degree of objectivity
\cite{altschuler2011,bornmann2010,bornmann2009,bornmann2008,campanario2007,wilson2006,bedleian2003,campanario1998a,campanario1998b,campanario1993,blank1991}.
The path from a request by a journal editor to review a certain
paper to the final submitted referee report is very similar to many
other activities, as for instance maintaining a correspondence. 
Here, we consider the task-completion of referees
 by looking at review processes
in journals Journal of High Energy Physics (JHEP) and Journal of
Statistical Mechanics: Theory and Experiment (JSTAT). We are
particularly interested in how the recommendation given by the
referee affects the review dynamics.

Clearly a manuscript to be reviewed presents an "important task" in the daily
life of a scientist. It takes place simultaneously with other daily
activities, including non-professional ones. It is susceptible to
a similar analysis recently undertaken for e-mail, text message, and
ordinary letter correspondence
\cite{barabasi05,malmgren08,barabasi05b,malmgren09,Wu10}, where
classical human activity models based on Poissonian statistics are
questioned in favor of power-law statistics
\cite{barabasi05,barabasi05b,jo2012,walraeyens2012}. The idea is that it might be so that human response dynamics do not have any particular time-scale.
Text message data \cite{Wu10} has been described by a combination of exponential and power-law distributions. This is motivated by an interplay of
Poisson-like initiation of tasks, decision making for task execution
and interactions among individuals that influence the dynamics. The
peer-review process has as such been analyzed on the level of
publication processing times, i.e. the time it takes for a
manuscript to appear from the initial submission
\cite{jo2011,mryglod11}.

Our analysis is based on two sets of data from the electronic-only
journals JHEP and JSTAT \cite{links},  with 7908 and 2558 review
requests leading to a referee report, respectively, each from
roughly a three-year period. Each entry contains the following
information about the review process: preprint id-number, preprint
submission date, id-number of the referee to whom the preprint is
assigned, date the preprint is assigned by editor, date the referee
accepted the assignment (if accepted), date the referee declined the
assignment (if declined), date the report was sent, version number
of the preprint and the editor's decision. The whole process is
illustrated in Figure
1a. If a preprint was assigned to several referees, each is
described with a separate entry. All dates are shifted with a random
interval to protect anonymity.

\section*{Results}
Figure 1b
depicts the distributions of the total waiting time
t(A)-t(E) (left) \cite{mryglod11,jo2011}) and t(B)-t(E) (right, time interval
between the assignment of a given manuscript to a given referee and
the date the referee report is sent) for both journals. We
immediately see that the idea of the referee response solely
determining the shape of waiting time distribution is not justified.
Fig. 1c shows again the distribution of t(A)-t(E),  in units where
time is rescaled with the mean waiting times. The both journals appear to have
similar characteristics.

A referee report is produced so that an incoming request is replied
to with a report, in contrast to correspondence where messages are
not always answered. Here, the referee gets reminders automatically;
21 (JSTAT) or 28 (JHEP) days after the assignment. The reports can
be classified indirectly using the resulting editor's decision.
There are a few possible outcomes: accepted, accepted with minor
corrections, to be revised, rejected, and not suitable for the
journal, denoted below by I, II, III, IV, and V, which last case
is quite rare and is in the following omitted from consideration.

One important question is whether the day-to-day variation has an
influence on the statistical properties (Fig. 2a).
A review typically takes much more time than a week, so considering the data
it appears that it does not matter when, during a week, such a request is made. This is in spite
of the fact that the request-statistics vary over a 7-day cycle.
When left free to allocate a given period of time between completing
easy and difficult tasks, it has been found in psychology that
people tend to allocate more time for the
easy ones \cite{payne2007}. Does a referee write the easier reports
(say for those manuscripts that are the very good and those clearly
not original or rigorous enough) faster? We split the data into ``immediate replies'' 
during the two first days, and the rest. Results in Fig. 2b
support this idea, since the fraction of both accepted and rejected
submissions is higher among reports from the first, fast case. This means that 
verdicts where the manuscript will need to be elaborated more due to the improvements
to be suggested are less frequent.

The waiting time statistics are found to be dependent on the version of the
manuscript (Fig. 2c).
For JSTAT, 43 \% of the manuscripts are sent after the editorial
decision to the authors for
revision whereas for JHEP the figure stands at 30 \%. The waiting
times for second or higher versions are not surprisingly generally
shorter. Fig. 2d depicts
the  distributions for various verdicts. 
The average waiting times (JSTAT and JHEP) are for the various cases:
I: 11.2 days; 15.0 days, II: 17.4 days; 22.0 days,
III: 19.8 days; 23.4 days, and IV: 17.1 days; 21.7 days. 
The data implies that JHEP has in general longer response times
than JSTAT. For the first round referee reports the differences are not very 
great among various cases. However, for later, revised versions, it is clear 
that reports which need to contain constructive criticism (accepted with minor revision, to be revised) are again much slower to do.
This feature is similar to that seen for the immediate replies.
JHEP and JSTAT have similar statistics after rescaling with the average report waiting time, and
as indicated  by the averages for any particular decision the associated
distribution has its specific character.  In particular the
distributions are not of a power-law, scale-free type. 

To account for the observations and the fact that people generally
tend to switch between tasks \cite{payne2007,gonzales2005} we construct and
apply an event-based model (Methods) based on multitasking: the work on a
manuscript is constantly mixed with competing activities. This is
formulated based on two kinds of competing tasks, both Poisson processes
with equal duration times. Task R is related to reviewing a
manuscript and task O means doing something else. Tasks are executed
until a sufficient amount of R-tasks is completed to finish the
review. A central assumption that we make is that the bias in the completion
times is manifest only in the number of tasks of any kind, whether O or R, to
be given attention to prior to finishing the report. To give attention does not
mean finishing the task at hand of course. We also take into account the in-built 
remainder system both journals use, inducing a Deadline. The details (Methods) mean,
that no refereeing-related (R) tasks are done before a certain time related to the
journal-induced timescale has passed.

Simulated data is shown with the original data separately for each
final verdict type and journal in Fig. 3.
The choice $\lambda = 96$ implies roughly a 15-minute average
duration for a single task (R/O), leaving two free parameters, $p_1$
and $p_2$, for each verdict-journal combination. For 
JSTAT and JHEP the four
cases I...IV can be grouped as pairs of values for $p_1$, $p_2$ for
the first and subsequent refereeing rounds as (0.0008, 0.0016),
(0.0014, 0.0004), (0.0012, 0.0004), and (0.001, 0.0008) and for
JSTAT as (0.0008, 0.0016), (0.0014, 0.0006), (0.0014, 0.0004), and
(0.001, 0.0008). To fit the model the four different cases have
similar parameters for \emph{similar decisions} for the two
journals. The different versions of manuscripts have quite different
waiting times - as noted since a referee has already once read the paper. 
The model does not include the natural circadian rhythm, so if we
discount one third of the total daily time for sleep, we arrive at a
10 minute average duration for a single task.

\section*{Discussion}
The typical values of $p_{i}$ indicate that the review process consists of
the order of a thousand subtasks, with a geometric distribution. Not
all of these steps imply refereeing: those steps are embedded in the
total activity such that a series of tasks ...OOOROOROO...R is
equivalent to ...OOOOOOROO....R. The single tasks to be completed
include reading the draft, writing the report etc. but these are
interrupted by tasks that demand execution (meetings, working on own
projects, shopping for groceries...).  For each manuscript the
number of tasks that need to be completed varies according to how
familiar referee is with the subject, how busy he/she is with
his/her own work, what is the general quality of the submission and
so forth - that is, it also is related to the verdict.

The dynamics of refereeing can be described by a geometric summation
scheme based on asymmetric Laplace law. Similar kinds of models find
applications \cite{Geometric} when studying phenomena of cyclic
nature. Both the data and the model indicate refereeing is described
by a "Deadline effect": the task has to be completed in the presence
of competing noise. Thus it is different from correspondence
patterns and the statistics of referee report completion
do not have scale-free features. Broad power-law distributions are excluded
by the need to complete the report. 

The contribution of a referee is a crucial factor determining the
duration of a review process. Fig. 1b shows that the shape of the
waiting time distribution is different if one considers the whole
review time compared to expectation from the time taken by the referee alone.
Therefore, the time taken by the editor to process a given
manuscript is also dependent on the "quality" of the paper. The 
statistics of the times it takes for the editor to send the manuscript
to a referee (t(B)-t(A)) correlate also with the final verdict, 
and are free of such a clear Deadline feature as seen 
in the referee statistics. Such waiting
times are probably in the case of JSTAT and JHEP a result - often - of 
lengthy searches of willing referees.
One should recall the common practice that authors suggest referees or
to exclude certain persons that they think might have a negative
prejudice about the work or authors. This has been indicated to have
a positive effect on getting published \cite{grimm2005}. Note that
the original selection of the referee may well be influenced by the
expectations of the editor, who thus have their important role
\cite{fischer2004}.

Considerations of waiting times and their origins
are related to the important issue of detecting fraud \cite{bornmann2008}.
One could for instance ask, what do the features we see here imply about
letting simply bad or even fraudulent manuscripts get published? 
Certainly our model includes a specific effect that is related to the
control mechanisms of JSTAT and JHEP, the reminders that are sent in other words.
 
The dynamics of the process are correlated with the final verdict.
In other words, it is easier to review a manuscript that seems at a
first glance "better", and this manifests as a bias in the
under-pressure situation of a continuous stream of decisions whether
to proceed with tasks related to the report. Note that the
microscopic time scale $\lambda$ is chosen in our simulations
similar for all O's and R's. The correlations between the measured statistics and the final
verdict present the question: is the determinism it implies a result
of a subjective referee bias, or that the expert referee follows a
justifed, educated guess from the beginning? We think that taken as
a trend, the causality implies the latter.

\section*{Methods} The decision model describes the properties of
tasks R and O and decides when the review process is finished. We
implement a  scheme which implies a geometric summation of
independent and identically distributed random variables, giving
rise to asymmetric Laplace
 distributions \cite{laplace}. We start with the assumption
 where both tasks R and O are identical Poisson
processes with an exponentially distributed duration time. Thus we
are left with a process where the random variable describing the
duration of the review process - for each verdict separately - is

\begin{equation}
 T_{w} = T'_{0} + \sum_{i=0}^{N_{j}} T_{i},
\end{equation}

\noindent where $T'_{0} \sim \mathrm{Uniform}(0,t_{0})$, $T_{i} \sim
\mathrm{Exponential}(\lambda)$ and $N_{j} \sim \mathrm{Geom}(p_{j})$
($j=1,2$, see below). The parameter $t_{0}$ for both journals was
interpreted to be the time interval after which a remainder for the
referee is automatically sent - 21 days for JSTAT, 28 days for JHEP
(independent best trial fits arrive surprisingly enough at close
values of $t_0$, data not shown).

For second and higher versions, we omit the $T'_{0}$ to make the
resulting distribution fit the exponential waiting time distribution
(presented in log-log-scale in Fig. 2c). Thus for both journals we
have model parameters $t_{0}$, $\lambda$, $p_{1}$ and $p_{2}$. To
test the model an event-driven simulation is run with the following
steps. i) Draw a uniform random number $t'_{0}$ with mean $t_{0}/2$,
ii) Draw a geometric random number $\nu_{1}$ with mean $1/p_{1}$.
iii) Draw and sum up with $t'_{0}$ $\nu_{1}$ exponential random
numbers with mean $1/\lambda$. iv) Draw a uniform random number $r$
between 0 and 1. v) If the value of r is smaller than f (f is the
fraction of manuscripts sent for revisions), draw a geometric random
number $\nu_{2}$ with mean $1/p_{2}$. If not, return to i). vi)
Continue summing with drawing $\nu_{2}$ exponential random numbers
with mean $1/\lambda$ and with $t'_0 =0$, Return to ii) and repeat
N-1 times. The sampling by simulations was chosen to be the same as
in the empirical data i.e. 7908 review events for JHEP and 2558 for
JSTAT, respectively. Each simulation was repeated 100 times and the
parameters $p_i$ were varied to obtain the best fit in each case.


\begin{thebibliography}{1}
\bibitem{links} jhep.sissa.it, jstat.sissa.it
\end{thebibliography}

{\bf Acknowledgemnts:} The Academy of Finland's Center of Excellence
-program is thanked for funding. The authors are grateful to the JSTAT and JHEP editorial office personnel and technical staff. Prof. Marc Mezar\'d should be thanked
for support as the JSTAT Chief Scientific Editor.

{\bf Author contributions:} TH analyzed the data and run the model simulations. MA had the main responsibility for manuscript writing.

{\bf Competing financial interests:} The authors delare no competing financial interests.

\begin{figure}[h!t!b!]
\includegraphics[width=0.7\textwidth]{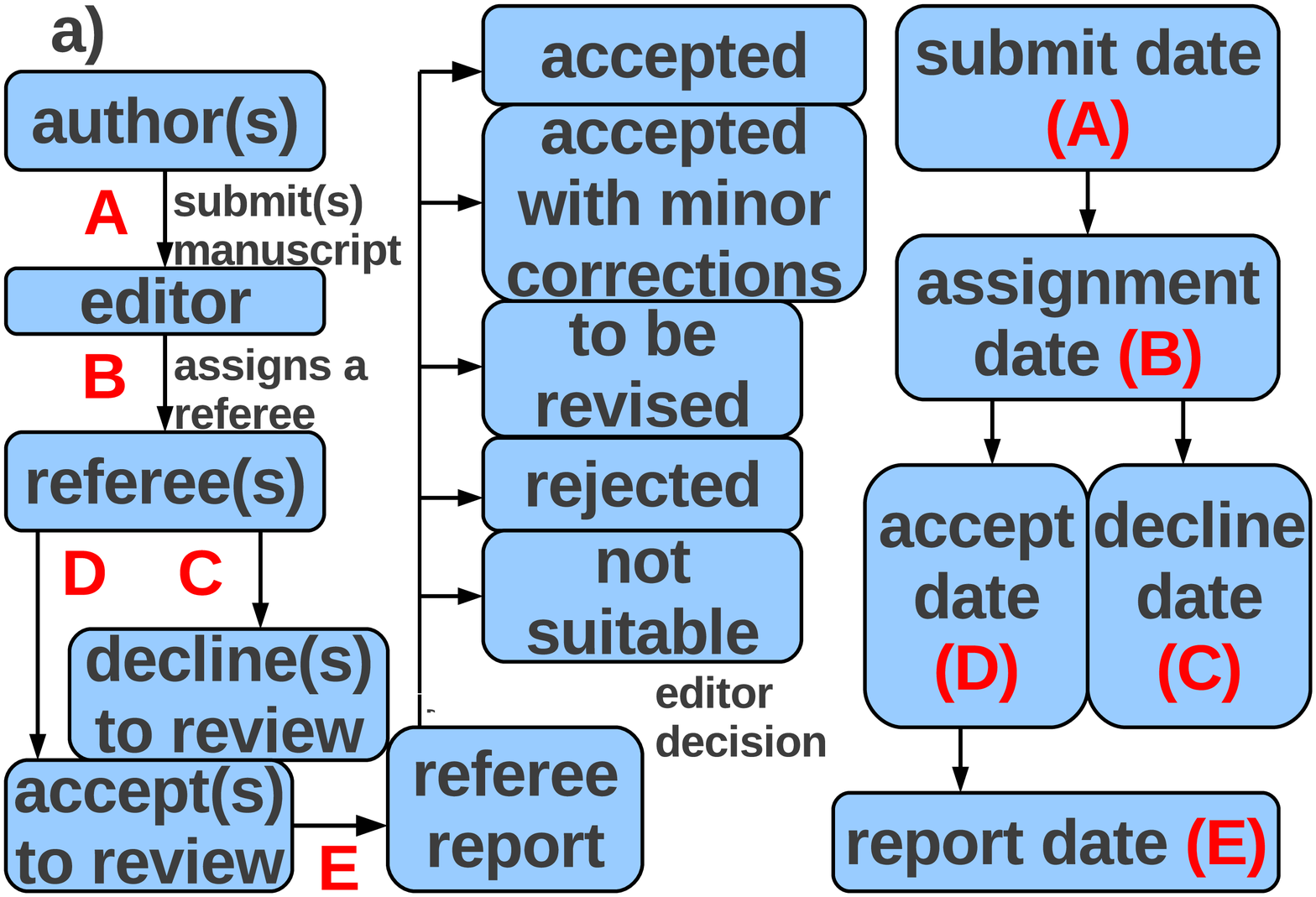}
\includegraphics[width=0.7\textwidth]{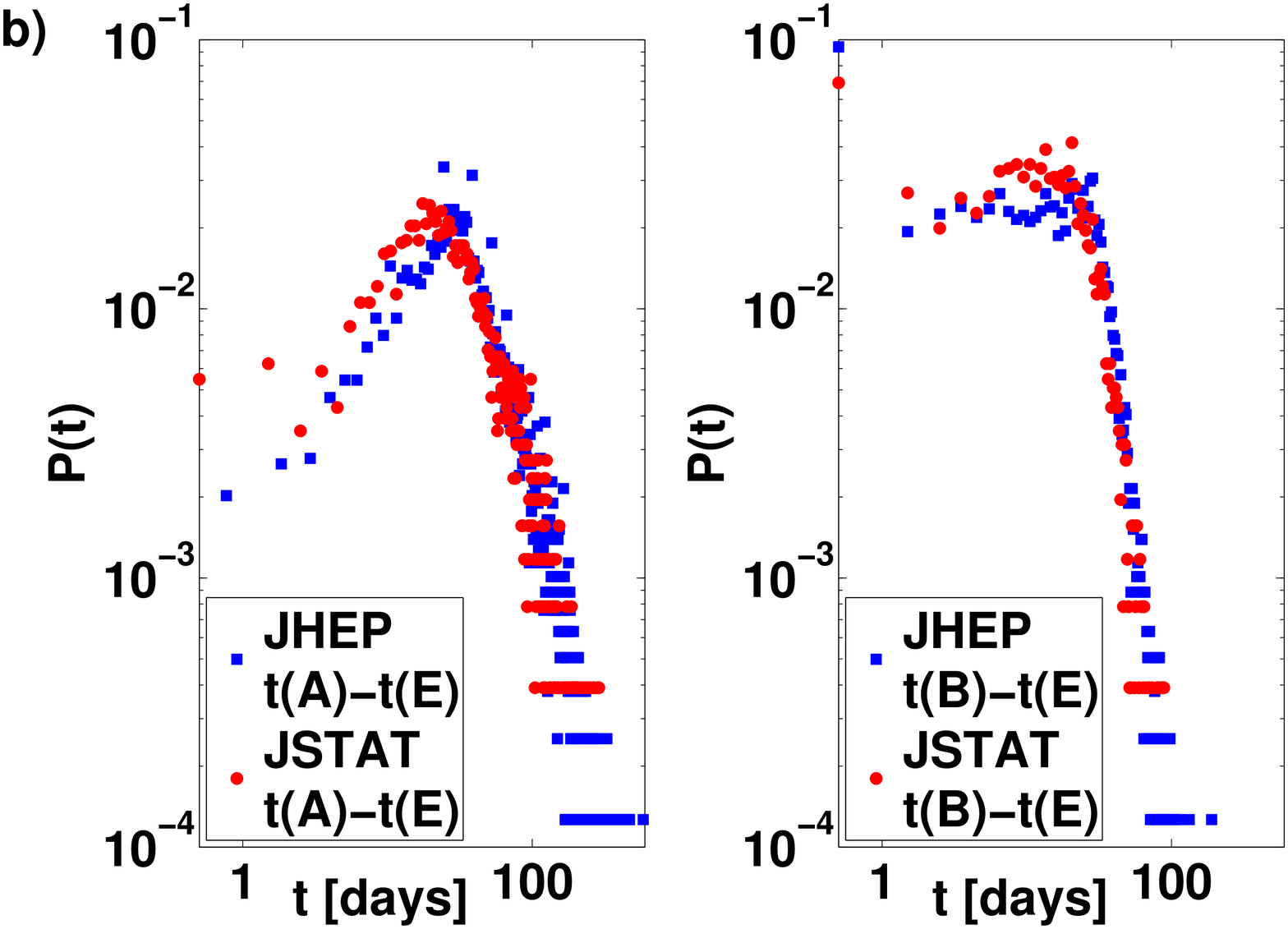}
\includegraphics[width=0.7\textwidth]{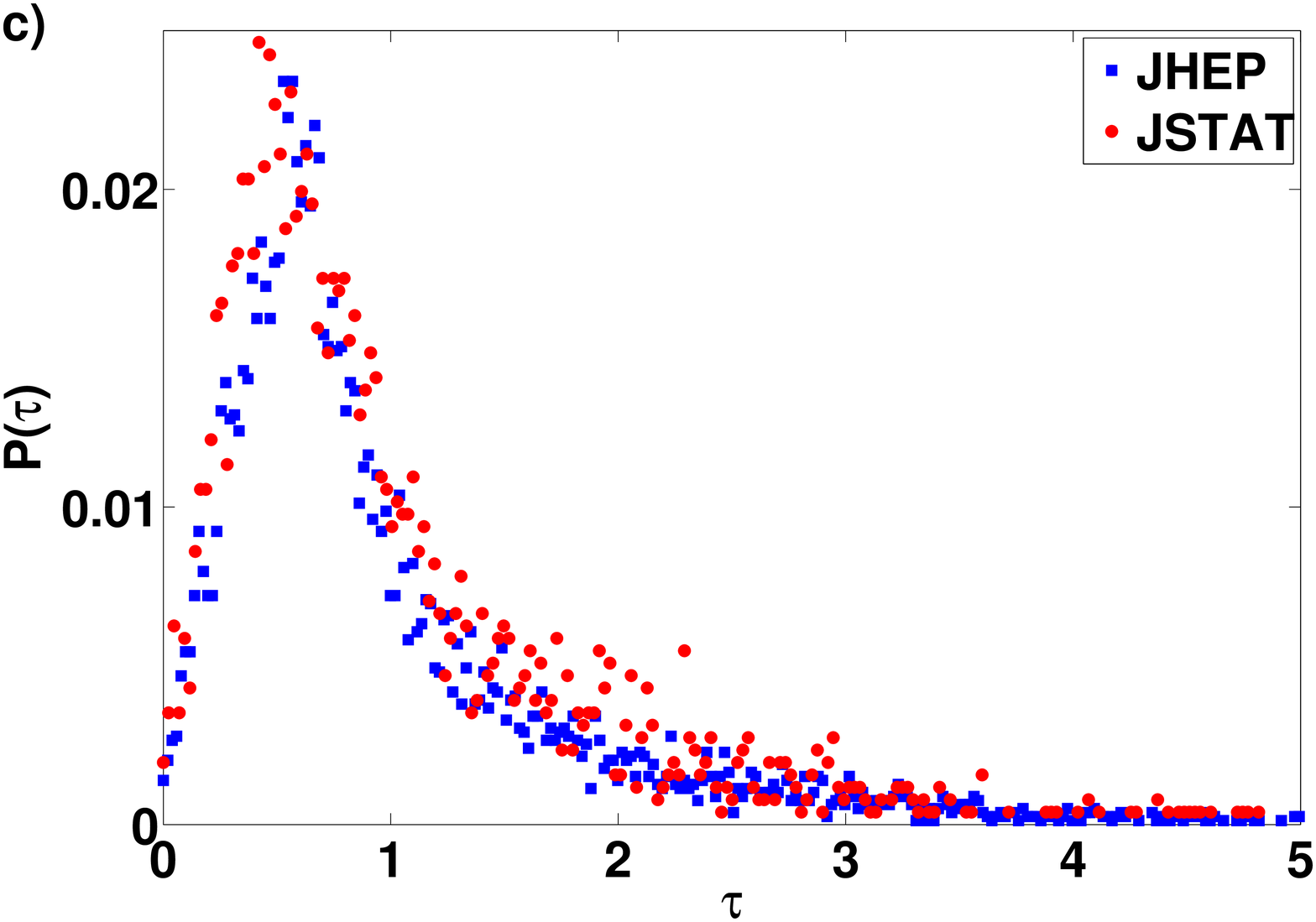}

\caption{\textbf{How to measure the processing of a manuscript.} a)
Editorial process in JSTAT and JHEP. b) Review process durations for
both journals. Left: $t(E)-t(A)$, right: $t(E)-t(B)$. c) Total
duration $\tau$ histogram of the review process $(t(E)-t(A)$) for both journals after
rescaling both distributions with their respective mean values.}
\label{fig:process}
\end{figure}

\begin{figure}[h!t!b!]
\centering
\begin{tabular}{c@{ }c@{}}
{\includegraphics[width=0.53\textwidth]{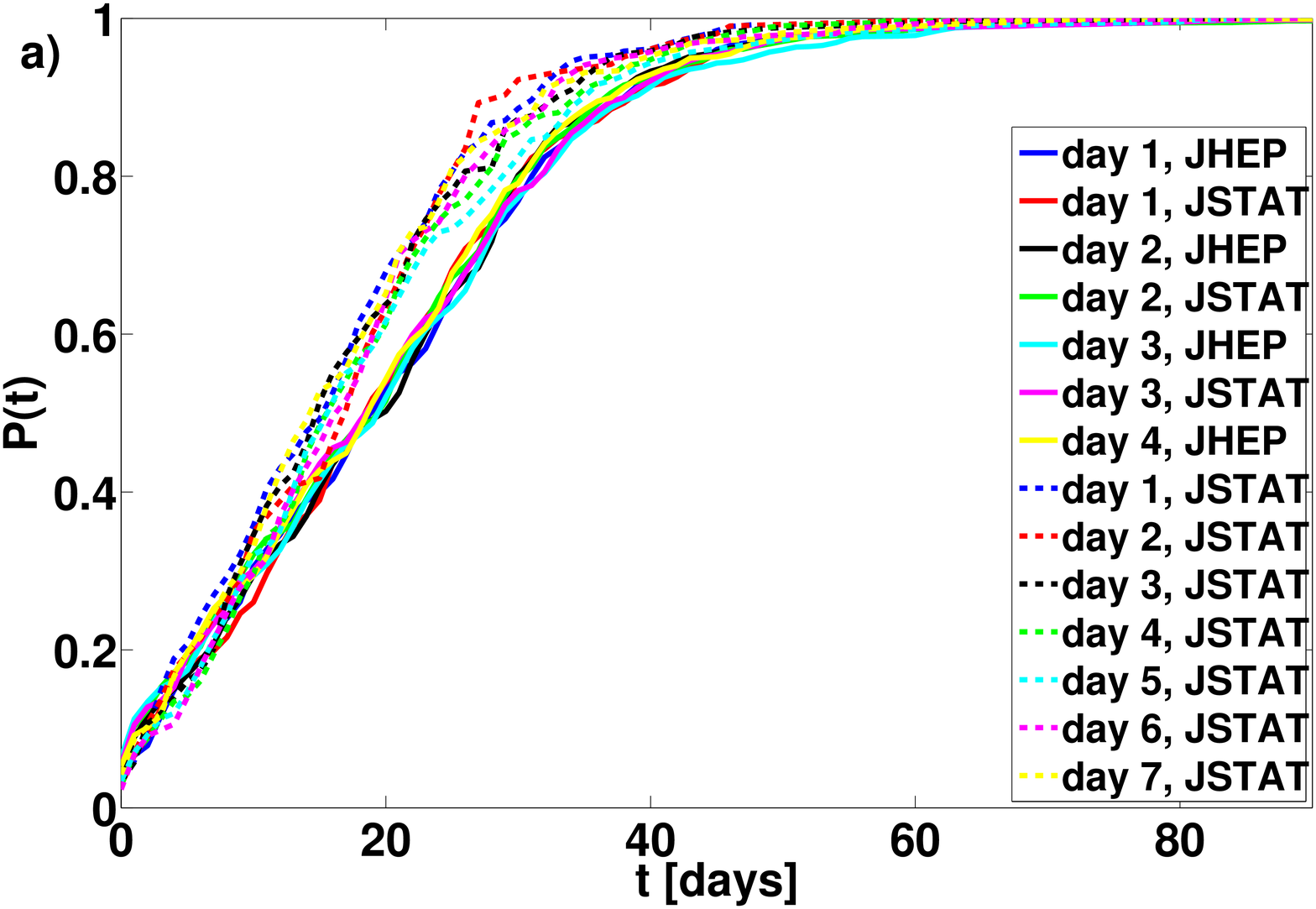}}
&{\includegraphics[width=0.55\textwidth]{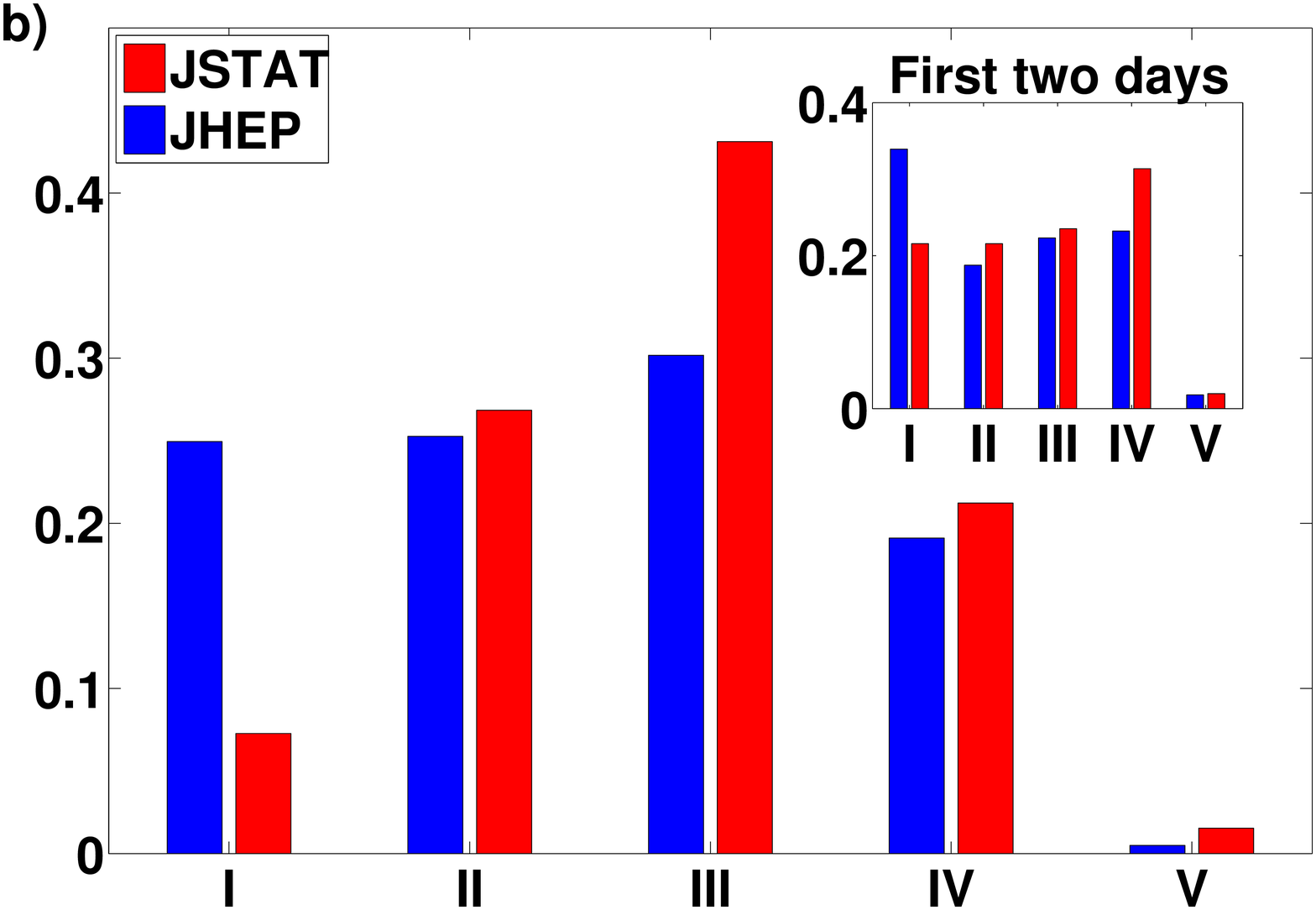}}
 \\
{\includegraphics[width=0.6\textwidth]{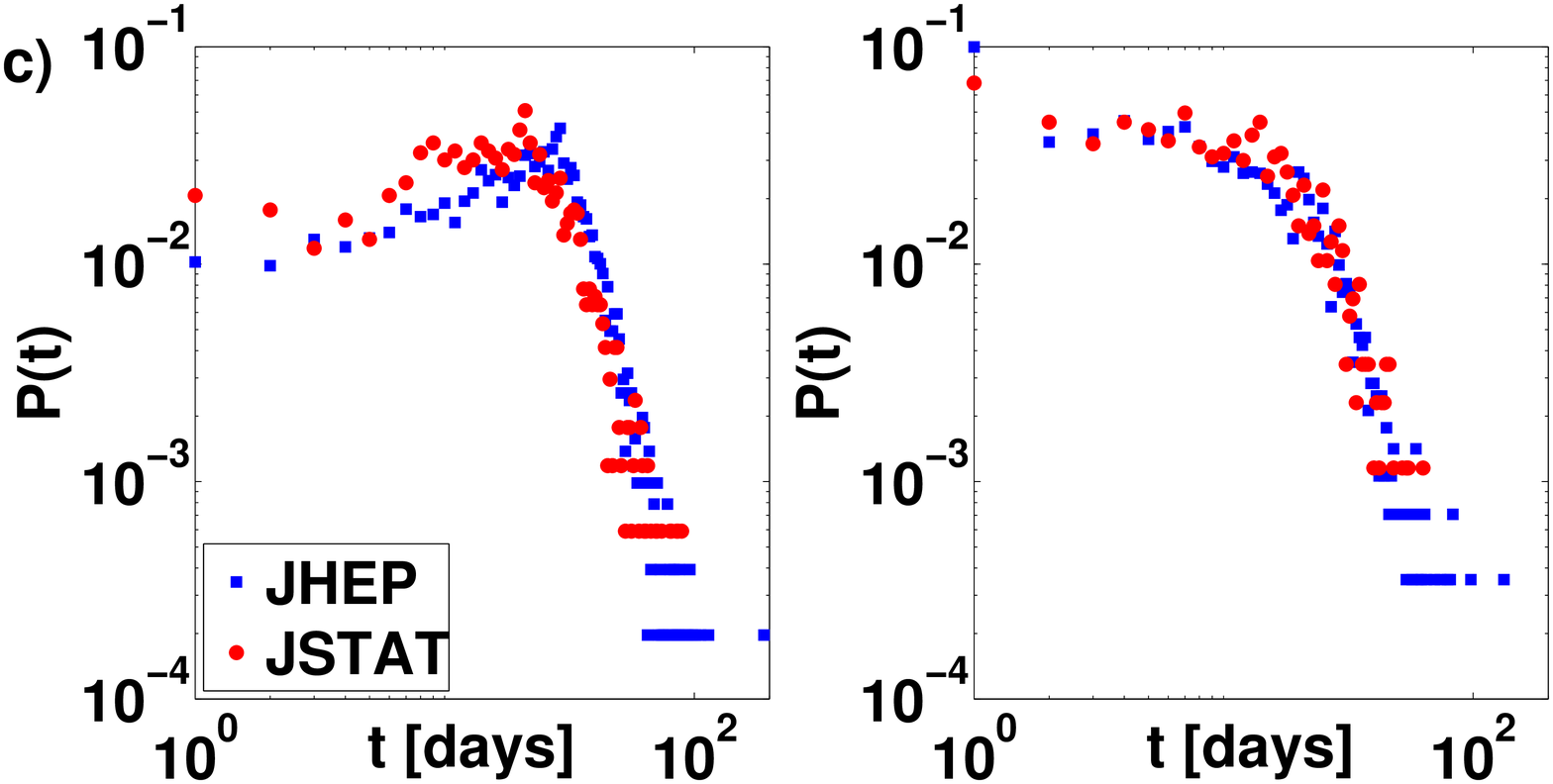}}\vspace{-4mm}
&{\includegraphics[width=0.6\textwidth]{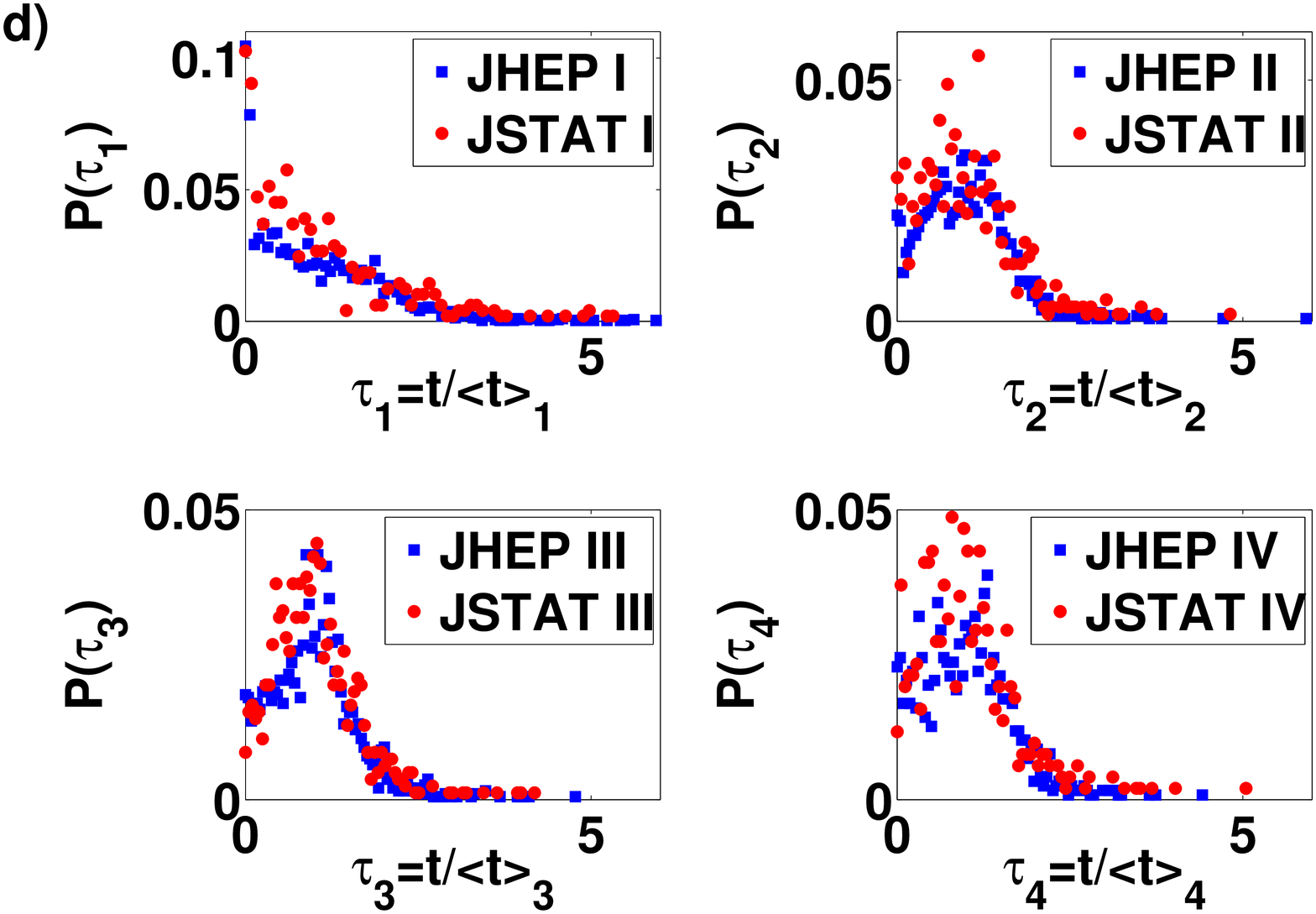}}
\\ 
\end{tabular}

\caption{\textbf{Data of the waiting times for referee reports.} a)
Weekday cyclicity (none): Cumulative distributions of waiting times
$t(A)-t(E)$ for manuscripts submitted on different weekdays for both
journals., b) Fractions of various decision: classes accepted (I),
accepted with minor revision (II), to be revised (III),  rejected (IV), not
appropriate (V) from all reports (as an inset the same for
short-duration processes) c) Waiting time distribution
$P(t(B)-t(E))$ for 1st versions of manuscripts (left) and higher
versions (right), d) Dependence of the review duration $(t(B)-t(E))$
on the verdict and journal. Distributions are scaled with the mean
durations.} \label{fig:data}
\end{figure}

\begin{figure}[h!t!b!]
\includegraphics[width=1.2\textwidth]{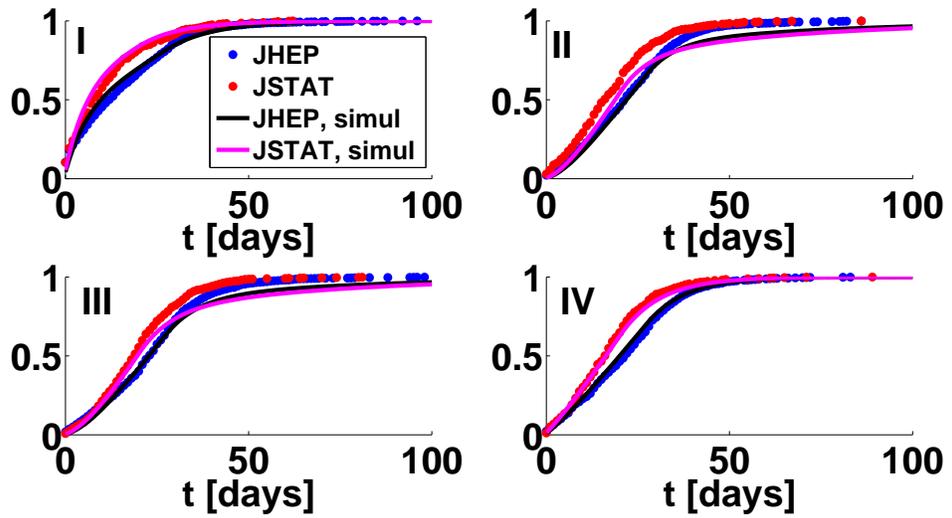}

\caption{\textbf{The event model against data. } We show the
aggregate cumulative distributions (for first and further rounds of
refereeing). The data is depicted for the four main categories
(accepted, accepted with minor revision, to be revised, rejected).
To compare with, the empirical first round average response times are (for JSTAT
and JHEP, respectively) 
(I: 23 days 12.5 h; 27 days 12 h), (II: 17 days 22.5 h; 22 days
10.5 h), (III: 21 days 10.5 h; 22 days 16.3 h), and (IV: 20 days 22 h; 24
days 10 h), respectively.} \label{fig:model}
\end{figure}


\end{document}